%% file: main.tex
\documentclass[manuscript,screen]{acmart}
\AtBeginDocument{%
  }

\begin{document}

\title{Designing for Constructive Civic Communication: A Framework for Human-AI Collaboration in Community Engagement Processes}

\renewcommand{\shorttitle}{Designing for Constructive Civic Communication}

\author{Cassandra Overney}
\email{coverney@mit.edu}
\affiliation{%
  \institution{MIT Center for Constructive Communication}
  \city{Cambridge}
  \state{Massachusetts}
  \country{USA}
}

\renewcommand{\shortauthors}{Overney}

\begin{abstract}
Community engagement processes form a critical foundation of democratic governance, yet frequently struggle with resource constraints, sensemaking challenges, and barriers to inclusive participation.
These processes rely on constructive communication between public leaders and community organizations characterized by understanding, trust, respect, legitimacy, and agency. 
As artificial intelligence (AI) technologies become increasingly integrated into civic contexts, they offer promising capabilities to streamline resource-intensive workflows, reveal new insights in community feedback, translate complex information into accessible formats, and facilitate reflection across social divides. 
However, these same systems risk undermining democratic processes through accuracy issues, transparency gaps, bias amplification, and threats to human agency. 
In this paper, we examine how human-AI collaboration might address these risks and transform civic communication dynamics by identifying key communication pathways and proposing design considerations that maintain a high level of control over decision-making for both public leaders and communities while leveraging computer automation.  
By thoughtfully integrating AI to amplify human connection and understanding while safeguarding agency, community engagement processes can utilize AI to promote more constructive communication in democratic governance.
\end{abstract}

\maketitle

\input{part1}
\input{part2}
\input{part3}
\input{conclusion}

\begin{acks}
The first author would like to express gratitude to her general exam committee, Deb Roy, Mahmood Jasim, and Mitchell Gordon, for their insightful reading recommendations and valuable feedback throughout the general exam process. 
She is also grateful to Pamela Siska at the MIT Writing and Communication Center for her thoughtful suggestions and editing assistance. 
Claude Sonnet was utilized for editing support and refinement of this manuscript, but all ideas presented in this essay are the author's own original work.
\end{acks}

\bibliographystyle{ACM-Reference-Format}
\bibliography{main}










\end{document}

%% file: part1.tex
\section{Communication Pathways in Community Engagement Processes}

This section first defines community engagement and then highlights two critical communication pathways: communication between public leaders and community organizations and communication between community organizations through public leaders.

\subsection{Defining Community Engagement Processes}

Community engagement processes are an important aspect of democratic governance and describe how governments at all levels involve the public in governing~\cite{corbettProblemCommunityEngagement2018}. 
Unlike civic engagement, where the public harnesses social capital to collectively address issues~\cite{putnam1995bowling}, community engagement specifically addresses ``municipality-to-citizen'' interactions and serves as a mechanism for sharing power with the public~\cite{roberts2015age, corbettProblemCommunityEngagement2018}. 
When implemented effectively, community engagement creates mutually beneficial outcomes for public officials and communities by strengthening decision-making quality and improving the adoption of policies~\cite{yavuzDigitalDivideCitizen2023}.

Community engagement occurs in various civic domains, including urban planning, policymaking, and public resource allocation~\cite{pengPathwayUrbanPlanning2023, nelimarkkaReviewResearchParticipation2019}.
The nature of this engagement varies considerably across projects but typically involves community members providing feedback to government entities. 
This feedback may be quantitative, such as votes or rankings, or qualitative, such as ideas, comments, and opinions from the public~\cite{mahyarCivicDataDeluge2019}.

Within community engagement processes, two primary stakeholder groups emerge: \textbf{public leaders} and \textbf{community organizations}. 
Public leaders encompass elected and appointed officials involved in civic decision-making~\cite{mahyarCivicDataDeluge2019}. 
The public is often organized through various community or civic institutions (e.g., housing authorities, community boards).
These stakeholders interact through several key communication pathways that facilitate the engagement process.

\subsection{Highlighting Key Civic Communication Pathways}

Communication between public leaders and community organizations forms a critical pathway that serves multiple purposes in civic decision-making. 
This bidirectional exchange helps bridge gaps in understanding, empathy, and agency between decision-makers and affected populations~\cite{cramerUsingForaPut}.
It also reduces relational distance between citizens and governments~\cite{corbettRemovingBarriersCreating2019, fungVarietiesParticipationComplex2006} while avoiding public disputes through two-way persuasion and education processes~\cite{irvinCitizenParticipationDecision2004}.

Communication among community organizations is equally important and occurs both with and without facilitation by public leaders.
Gastil highlights the significance of deliberation and dialogue among citizens to create a ``deliberative community'' in political communication~\cite{gastil2008political}.
Providing opportunities for community organizations to form and maintain relationships with each other can strengthen existing community assets and result in sustainable community engagement outcomes~\cite{dickinsonCavalryAinComing2019}.
Nelimarkka suggests that technology-enabled participation should support participants' efforts to develop common ground and enable collaboration among diverse groups~\cite{nelimarkkaReviewResearchParticipation2019}.   
This communication pathway allows for bottom-up organization and agenda-setting that complements traditional top-down governance approaches. 

\begin{figure}[h]
  \centering
  \includegraphics[width=\linewidth]{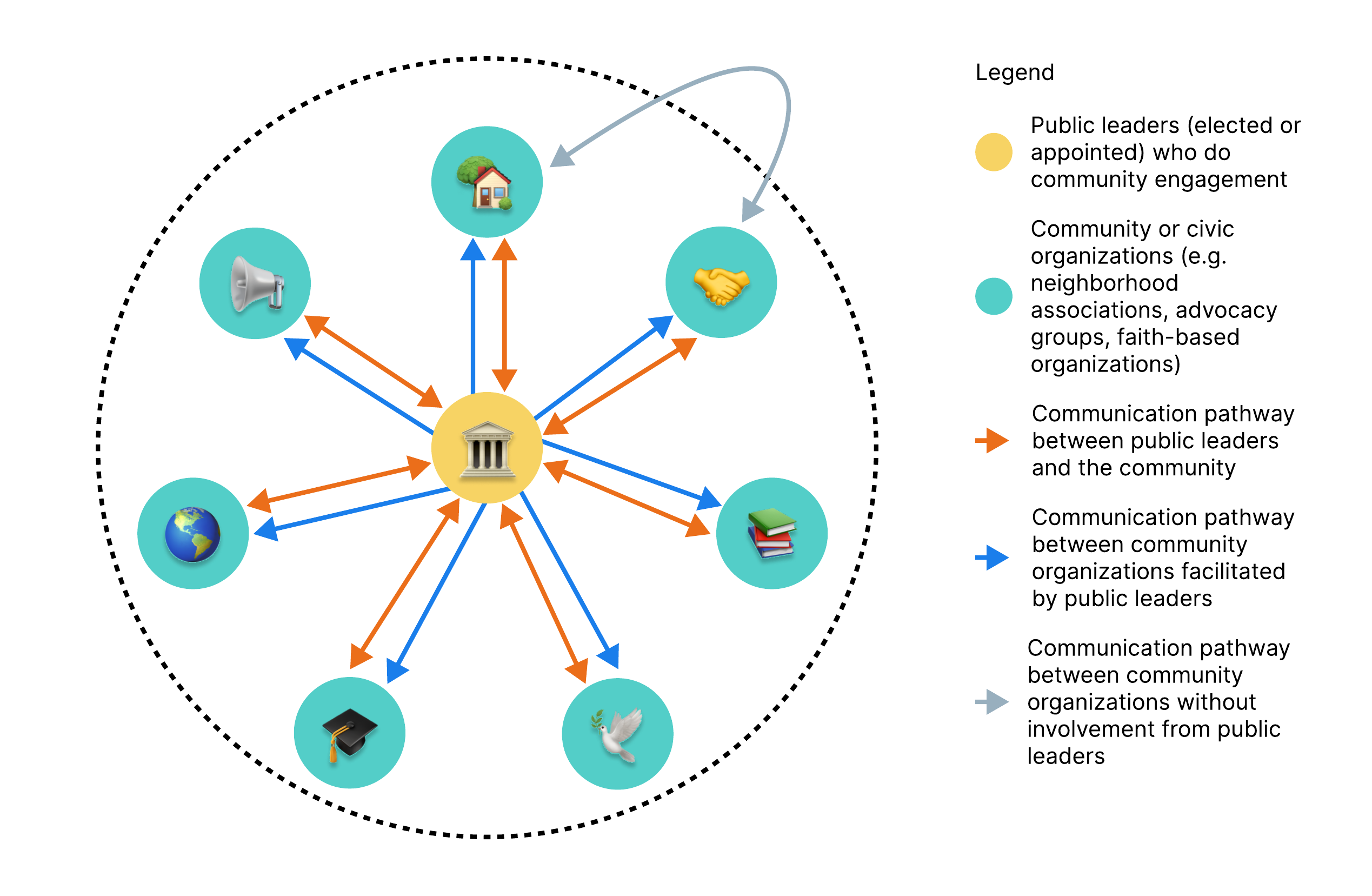}
  \caption{A diagram of the key communication pathways in a typical community engagement process.}
  \label{fig:figure1}
\end{figure}

Fig.~\ref{fig:figure1} displays a diagram of the communication pathways. 
We focus primarily on two communication pathways involving public leaders who initiate community engagement efforts: the bidirectional communication between public leaders and community organizations (orange lines), and the communication between community organizations facilitated by public leaders (blue lines).
These pathways form the foundation for constructive community engagement.


%% file: part2.tex
\vspace{-0.5cm}
\section{Defining Constructive Civic Communication and Existing Challenges}
This section describes how to achieve constructive communication in community engagement processes and what challenges hinder this goal.
We explore constructiveness for two individual communication pathways as well as the overall system presented in Fig.~\ref{fig:figure1}.

\subsection{Communication between Public Leaders and Community Organizations}

Constructive communication between public leaders and community organizations can be assessed through several outcome measures. 
For public leaders, \textbf{understanding} represents increased comprehension of community perspectives and potential knowledge gains from community input.
Additionally, \textbf{trust} in community input and \textbf{respect} for community expertise are critical indicators of constructive communication. 
For community members, critical outcomes include increased \textbf{understanding} of governance processes, \textbf{trust} in leaders' engagement capabilities, and a sense of \textbf{legitimacy} regarding the participation process. 
Most importantly, community members experience \textbf{agency} when they feel their participation matters, regardless of their alignment with final outcomes.

Public leaders face significant barriers to achieving constructive communication with community organizations.
First, designing effective community engagement processes requires substantial resources and expertise~\cite{irvinCitizenParticipationDecision2004, brysonDesigningPublicParticipation2013}.
Community engagement encompasses a heterogeneous collection of methods and goals that vary widely across governments, leading to uneven implementation and inconsistent outcomes~\cite{corbettProblemCommunityEngagement2018}. 
The design of public participation processes represents a complex endeavor requiring careful consideration of multiple stakeholder needs~\cite{brysonDesigningPublicParticipation2013}. 
Public leaders report significant difficulties in formulating questions that elicit actionable information from deliberation at scale~\cite{smallPolisScalingDeliberation2021, overney2025coalesce}. 
Even after collecting community input, preparing comprehensive and accessible responses demands considerable effort.
Officials must synthesize extensive feedback, construct coherent narratives about lengthy engagement processes, justify decisions, identify and address misconceptions, and transform raw data into accessible formats. 
These challenges frequently result in either no response to community input, responses that are overwhelming, or significant time delays that undermine community trust in the process.

Sensemaking of large-scale qualitative feedback presents significant challenges as public leaders confront overwhelming amounts of multifaceted and unstructured data in various formats~\cite{reynanteFrameworkOpenCivic2021} that often remains unutilized in policymaking~\cite{jasimCommunityPulseFacilitatingCommunity2021}.
Analysis processes are costly~\cite{mahyarCivicDataDeluge2019} and complicated by multiple biases from vocal minorities, analysts' personal perspectives, and methodologies that oversimplify marginalized viewpoints~\cite{baumerCourseItPolitical2022}.
Additionally, managing conflicting stakeholder perspectives becomes particularly challenging when addressing politically contentious issues or navigating contradictory constraints imposed by various social regulations and laws~\cite{ schattschneider1960contagiousness, carpenter2001public}.
A fundamental challenge remains in how decision-makers can appropriately weigh different inputs when individual opinions are not equal and must be understood within their social context~\cite{schumpeterCapitalismSocialismDemocracy2010, blumer1948public}.
From the community perspective, engagement fatigue represents a significant barrier when community organizations are repeatedly asked similar questions without evidence of institutional memory~\cite{taylorViewpointEmpoweringCommunities2012}.
This engagement fatigue is sometimes compounded by an overload of information from public leaders, creating a cycle where communities become increasingly reluctant to participate in engagement efforts~\cite{overney2025boundarease}.

More critically, a lack of response or inaccessible responses undermine constructive communication between community organizations and public leaders.
Community members frequently report sending feedback ``into a void'' without understanding how their input influenced decisions~\cite{jasimCommunityClickCapturingReporting2021}.
This lack of communication leads to engagement fatigue and distrust among participants about whether their opinions will prompt action.
When responses from public leaders exist, they often take inaccessible forms, such as long PDF reports with little interactivity, or information scattered across multiple platforms~\cite{overney2025boundarease}.
The general public is rarely considered when designing responses, resulting in responses that lack personalization, multiple entry points, or the ability to explore decisions interactively~\cite{baumerCourseItPolitical2022}.
Further complicating matters, public leaders sometimes hesitate to communicate summaries back to the public due to concerns about representativeness of the data~\cite{mahyarCivicDataDeluge2019}, essentially undermining the entire purpose of the engagement process. 
Without accessible, transparent responses that acknowledge community input, the communication pathway becomes fundamentally broken, regardless of how well-designed the initial engagement process might have been.

Underlying the communication challenges are relationship issues between public leaders and community organizations. 
Expectation mismatches occur when community members assume public leaders possess greater authority to implement changes than they actually have~\cite{irvinCitizenParticipationDecision2004}.
Additionally, a tension exists between transactional and relational interactions.
While technology often aims to make community engagement more efficient, this can conflict with relationship-building needs, in which relationships take time to develop.
The gap between the transactional engagements of customer service and the relational engagements of ``deep hanging out'' represents a central challenge in civic communication that technology must navigate carefully~\cite{corbettProblemCommunityEngagement2018}.

Addressing the communication challenges between public leaders and community organizations requires thoughtful design interventions, such as systems that enhance the user experience of designing and implementing community engagement processes.
The complexity and resource intensity of effective participation design can be mitigated through systems like Coalesce, which helps public leaders formulate questions that elicit actionable information~\cite{overney2025coalesce}.
To mitigate sensemaking challenges, tools like CommunityPulse~\cite{jasimCommunityPulseFacilitatingCommunity2021}, SenseMate~\cite{overneySenseMateAccessibleBeginnerFriendly2024}, and the Feedback Map Tool~\cite{beefermanFeedbackMap2023} provide structured approaches to analyze qualitative input.
These systems aim to reduce the resource burden on public leaders while increasing the quality of engagement.

Transparent responses address the void community members report when providing feedback~\cite{jasimCommunityClickCapturingReporting2021}. 
Effective systems justify decisions by showing real-time results~\cite{taylorViewpointEmpoweringCommunities2012, valkanovaMyPositionSparkingCivic2014a, smallPolisScalingDeliberation2021} and enabling exploration through accessible data visualizations~\cite{regan2015designing, faridani2010opinion} and personalized interfaces~\cite{overney2025boundarease}.
Implementing different levels of granularity through interactive visualizations can address public leaders' concerns around sharing unrepresentative summaries~\cite{mahyarCivicDataDeluge2019} by adding nuance and transparency to the decision-making process.

Constructive communication also depends on establishing feedback loops that change throughout the policy-making process~\cite{yavuzDigitalDivideCitizen2023}.
Multiple entry points for long processes should be provided, applying principles from design thinking frameworks to create iterative and dynamic engagement~\cite{reynanteFrameworkOpenCivic2021}. 
As Bryson et al. note, stakeholder engagement may evolve over the course of a participation process, and the purpose itself might change, making an iterative approach with clear feedback loops essential for allowing necessary adaptations to occur~\cite{brysonDesigningPublicParticipation2013}.
However, regardless of how many iterations exist in an engagement process, achieving constructive communication depends on ensuring community organizations have meaningful power and authority.
As Arnstein argues, participation without redistribution of power is a pointless and frustrating process for the powerless. 
When communities lack sufficient power, engagement remains at lower participation levels that maintain status quo power dynamics~\cite{arnsteinLadderCitizenParticipation1969}. 
By exploring these design considerations, civic technologies can help transform the communication channel between public leaders and community organizations into one characterized by mutual understanding and trust.

\subsection{Communication between Community Organizations Facilitated by Public Leaders}


Several outcome measures indicate constructive engagement between community organizations, especially across social and political divides.
\textbf{Understanding}, or horizontal transparency, represents increased comprehension of others' perspectives within the community, particularly those from different backgrounds~\cite{stainsjr.ReflectionConnectionDeepening2012, aitamurtoFiveDesignPrinciples2015}.
\textbf{Trust} emerges when participants believe in the reliability and authenticity of others' experiences~\cite{hagmannPersonalNarrativesBuild2020}, and \textbf{respect} develops as community members acknowledge the humanity in others~\cite{kubinPersonalExperiencesBridge2021, carpenter2001public, sternstrategygroup:speaking&advisoryandprDebRoyTrust2021}. 
Beyond a single interaction, constructive communication fosters \textbf{willingness to engage further} and even advocate for others whose needs differ from one's own~\cite{kubinPersonalExperiencesBridge2021, gastil2008political}.
Though this communication pathway primarily serves community members, public leaders also benefit when community feedback becomes more nuanced and comprehensive.

The diversity of motivations and expertise among community members presents a substantial challenge to constructive communication.
Many people are reluctant to participate without confidence that their input will translate into tangible action~\cite{fungVarietiesParticipationComplex2006, irvinCitizenParticipationDecision2004}.
Others lack the capacity—time, energy, or technical knowledge—to engage meaningfully~\cite{reynanteFrameworkOpenCivic2021}. 
Schumpeter observed a larger phenomena where communities exhibit a reduced power of discerning facts, a reduced willingness to act, and a reduced sense of responsibility, which amplifies complacency in community engagement~\cite{schumpeterCapitalismSocialismDemocracy2010}.
Addressing these motivation and expertise barriers requires careful design that acknowledges diverse capacities while creating accessible entry points.


Further complicating effective communication are echo chambers and ideological bubbles that limit exposure to diverse perspectives. 
The ``sense of fracture'' in contemporary civic discourse~\cite{sternstrategygroup:speaking&advisoryandprDebRoyTrust2021} creates fertile ground for public disputes~\cite{carpenter2001public}, which can derail constructive communication. 
When these divisions interact with digital participation technologies, technological interventions can amplify already dominant voices. 
The digital divide—encompassing access to technology, technical skills, and offline impacts of technology—can exclude marginalized communities from digital civic processes~\cite{yavuzDigitalDivideCitizen2023}, potentially reinforcing existing power imbalances through practices like super-posting by already-engaged community members.

Designing for constructive communication between community organizations requires enhancing the user experience of participation, especially for disadvantaged populations~\cite{yavuzDigitalDivideCitizen2023}.
Creating accessible and simple ways for the public to engage are essential~\cite{schattschneider1969two}.
Various systems demonstrate promising approaches to explain complex decisions, from visualizations showing boundary changes~\cite{overney2025boundarease} to tools for budgeting and planning~\cite{kimFactfulEngagingTaxpayers2015, tianParticipatoryEplanningModel2023}. 
Participation is made more accessible through multiple interfaces including chat~\cite{jiangCommunityBotsCreatingEvaluating2023}, mobile prompts~\cite{graeffActionPathLocationbased2014}, and situated civic technologies~\cite{schroeterEngagingNewDigital2012, koemanEveryoneTalkingIt2015, fredericksDigitalPopUpInvestigating2015}.
Crowdsourcing, or the creation of micro-tasks, provides another source of inspiration when designing systems that leverage the varying expertise and motivations among community organizations~\cite{reynanteFrameworkOpenCivic2021, mahyarCommunityCritInvitingPublic2018, ledantec2015planning}.

In addition, the nature of content exchanged between community organizations determines communication quality.
Personal experiences rather than abstract opinions foster more productive dialogue across divides~\cite{sternstrategygroup:speaking&advisoryandprDebRoyTrust2021, blackDeliberationStorytellingDialogic2008}. 
Kubin et al. found that issue-relevant personal stories involving harm increase perceived rationality and respect across moral divides~\cite{kubinPersonalExperiencesBridge2021}, while Kalla and Broockman demonstrated that perspective-getting through narratives reduces exclusionary attitudes~\cite{kallaWhichNarrativeStrategies2023}. 
When facilitated properly, sharing experiences helps communities focus on underlying interests rather than positions~\cite{carpenter2001public}, which can mitigate conflicts. 
Some platforms enhance this human connection through audio, as hearing someone's voice humanizes them more effectively than text alone~\cite{schroeder2017humanizing}. 
By thoughtfully designing for these content characteristics, public leaders can transform potentially divisive exchanges into opportunities for genuine understanding and respect across community differences.

Another design consideration includes determining the appropriate distance between community members, especially during cross-cutting exposures that bridge echo chambers~\cite{mutz2006benefits}.
Kriplean et al. proposes ``augmented personal deliberation'' which describes engaging with other perspectives at a distance~\cite{kripleanSupportingReflectivePublic2012} either through perspective-taking~\cite{kimCrowdsourcingPerspectivesPublic2019, bowenMetroFutures20202023} or sampling algorithms that expose users to feedback from individuals with different stances or from varying segments of a population~\cite{nelimarkkaComparingThreeOnline2014, kripleanThisWhatYou2012}.
Augmented personal deliberation allows people to access different thoughts without direct contact to others, which can mitigate the activation of political identity and flaming.
In contrast, there are some instances when allowing community members to be in direct contact with each other is beneficial, requiring the design of public deliberation systems~\cite{aitamurtoFiveDesignPrinciples2015, kleinHowHarvestCollective2007}.
Drawing upon dialogue practices~\cite{palmer2009hidden, blackDeliberationStorytellingDialogic2008, bohm2004dialogue, stainsjr.ReflectionConnectionDeepening2012}, which emphasize facilitation and prompts that elicit personal experiences and reflection, when designing deliberation systems can offer alternatives to social media patterns that often privilege speed and reactivity over thoughtful engagement~\cite{sternstrategygroup:speaking&advisoryandprDebRoyTrust2021}.
The spectrum of augmented personal deliberation to public deliberation relates to the ``intimacy gradient'', which emphasizes the agency for people to choose what level of exposure to others they desire~\cite{alexander1977pattern}.
Effective civic platforms should offer carefully designed modalities through which users can navigate according to their comfort level, similar to the architectural concept of ``circulation realms''~\cite{alexander1977pattern}.  
Even in a top-down process like community engagement, providing community organizations with bottom-up agency to participate is critical to promote constructive communication. 

\subsection{Overall System Design Considerations for Civic Communication}

Beyond examining individual communication channels, public leaders must consider how these pathways function together as an integrated system.
One consideration is the inclusiveness of communication opportunities across different channels.
Traditional engagement methods rarely provide equitable access, resulting in uneven representation where only a minority of privileged individuals participate~\cite{jasimCommunityClickCapturingReporting2021, mahyarCivicDataDeluge2019}.
This selective participation becomes ``a gateway for participation'', preserving power among a small number of community organizations~\cite{corbettProblemCommunityEngagement2018}.
The consequences extend beyond representation.
Inaccessible governance structures create a lack of accountability and transparency, which lead to distrust towards public leaders~\cite{dickinsonCavalryAinComing2019}.
Designing for inclusiveness requires addressing technological and language barriers~\cite{aitamurtoFiveDesignPrinciples2015} and actively considering the needs of disadvantaged populations~\cite{yavuzDigitalDivideCitizen2023}.

While the proliferation of multiple communication channels might appear beneficial, the volume and scale of communication presents its own challenges.
Several community engagement frameworks recommend implementing multiple channels of public participation~\cite{brysonDesigningPublicParticipation2013, reynanteFrameworkOpenCivic2021}. 
However, Schumacher's caution against the dangers of "giantism" suggests that scaling in small, manageable ways may prove more effective than maximizing the number of channels~\cite{schumacher1973small}.
This approach recognizes that increasing communication volume without corresponding increases in administrative capacity can overwhelm both public leaders and community organizations, potentially undermining meaningful engagement. 
Rather than creating numerous shallow engagement opportunities, constructive communication systems might benefit from fewer, deeper channels that support more dialogue while remaining accessible to diverse participants.


%% file: part3.tex
\section{Considering the Role of AI in Civic Communication}

The challenges in community engagement processes point toward a need for innovative approaches. 
Artificial intelligence (AI) presents a compelling opportunity to streamline workflows, reveal overlooked patterns, enhance stakeholder understanding, and create accessible engagement channels. 
However, introducing AI into civic communication brings a unique set of benefits, risks, and considerations that must be critically examined.

\subsection{Benefits of AI in Civic Communication}

AI systems can significantly streamline resource-intensive community engagement processes like sensemaking~\cite{zhangRedefiningQualitativeAnalysis2023}.
By automating routine aspects of engagement design and data processing, AI could enable public leaders to redirect resources toward relationship-building.
Simplified workflows can increase both efficiency~\cite{choi2024proxona} and performance with reduced effort~\cite{shneidermanHumanCenteredArtificialIntelligence2020}, making comprehensive community engagement feasible even for resource-constrained public leaders.
Furthermore, by making engagement processes less burdensome and more systematized, AI can help maintain institutional memory and support more consistent implementation across projects~\cite{corbettProblemCommunityEngagement2018}.

AI can enhance mutual understanding by translating complex information into accessible formats. 
AI-enhanced visualization and explanation tools enable meaningful exchanges by supporting human mastery of complex information, empowering users with varying expertise levels to engage more effectively in civic dialogue~\cite{shneidermanHumanCenteredArtificialIntelligence2020}.

AI can reveal patterns in community input that might otherwise remain obscured, helping identify diverse insights from unstructured qualitative data~\cite{mahyarCivicDataDeluge2019, zhangRedefiningQualitativeAnalysis2023} and highlighting commonalities that inspire cross-cutting exposures among community organizations~\cite{smallOpportunitiesRisksLLMs2023}. 
By presenting alternative perspectives in non-threatening ways, AI can foster reflection and facilitate collective growth~\cite{caiAntagonisticAI2024}.
Additionally, AI can serve as a conduit for spreading effective practices across different contexts, strengthening and diversifying ideas around how to design and implement engagement processes.

\subsection{Risks of AI in Civic Communication}

While AI offers a variety of potential benefits for civic communication, it also presents substantial risks.
The complexity of community engagement processes, with their reliance on trust, context, and understanding, makes them particularly vulnerable to AI's limitations.

AI systems can struggle with accuracy and reliability issues.
Hallucinations, which occur when AI generates false information~\cite{huang2023survey, zhang2023siren}, poses significant risks when applied to civic contexts~\cite{smallOpportunitiesRisksLLMs2023}.
In addition, AI often demonstrates limited understanding of local contexts~\cite{pengPathwayUrbanPlanning2023}, potentially overlooking neighborhood-specific concerns or culturally significant factors.
Regarding reliability, AI outputs remain highly dependent on input quality~\cite{fishGenerativeSocialChoice2023}, creating vulnerability in processes where community input may be highly varied and ambiguous.

The opacity of AI decision-making presents a challenge to democratic processes, which depend on transparency.
When applied to civic communication, AI's ``black box'' nature creates outputs that are difficult to verify or explain~\cite{pengPathwayUrbanPlanning2023, zhangRedefiningQualitativeAnalysis2023, smallOpportunitiesRisksLLMs2023}, threatening the legitimacy of engagement processes.
This creates what Fish et al. describe as a ``legitimacy-scalability tradeoff curve'' where increased efficiency often comes at the cost of decreased transparency~\cite{fishGenerativeSocialChoice2023}. 
This tradeoff is particularly problematic in civic contexts where perceived legitimacy directly influences community trust and participation. 
Without explicit disclosure of when and how AI is involved in communication processes, community members may question the authenticity of responses from public leaders, potentially deteriorating trust and respect between stakeholders.

In addition, AI systems risk amplifying existing inequities within civic processes through multiple forms of bias.
Data biases emerge when training datasets under-represent marginalized communities~\cite{pengPathwayUrbanPlanning2023, zhangRedefiningQualitativeAnalysis2023, smallOpportunitiesRisksLLMs2023} or include perspectives deemed socially unacceptable~\cite{leeWeBuildAIParticipatoryFramework2019}.
Model biases compound the fairness problem, with large language models promoting ecological fallacies~\cite{gordonJuryLearningIntegrating2022} and demonstrating biases against specific demographic groups~\cite{fishGenerativeSocialChoice2023}, which can introduce new forms of discrimination~\cite{burtonAlgorithmAversionHumanMachine2023, binnsFairnessMachineLearning2018}.
These biases particularly threaten inclusivity in community engagement by potentially reinforcing uneven representation of community perspectives.
For public leaders attempting to make sense of diverse community input, AI-produced summaries may appear comprehensive while actually failing to capture marginalized viewpoints, creating a false consensus that undermines the goal of creating inclusive public engagement opportunities~\cite{smallOpportunitiesRisksLLMs2023, slattery2024systematic}.  

AI in civic communication can suffer from both overreliance and underreliance. 
Overreliance occurs when users delegate critical decisions without appropriate oversight~\cite{burtonAlgorithmAversionHumanMachine2023}, creating dependency relationships where humans process AI information superficially~\cite{gajosPeopleEngageCognitively2022}.
Conversely, underreliance happens when stakeholders categorically reject algorithmic assistance~\cite{burtonAlgorithmAversionHumanMachine2023}. 
Finding the appropriate balance between overreliance and underreliance is critical for effective AI support systems.

More broadly, AI in civic contexts could create ``footloose technology'' where systems operate at scales beyond human control ~\cite{schumacher1973small}, potentially de-skilling stakeholders ~\cite{shneidermanHumanCenteredArtificialIntelligence2020} and diminishing appreciation for human capabilities~\cite{slattery2024systematic}. 
Simultaneously, technical vulnerabilities, including privacy breaches and security exploits like prompt injection~\cite{smallOpportunitiesRisksLLMs2023}, threaten the trust essential for authentic engagement. 
These risks necessitate human-AI collaborative systems that maintain meaningful human control while enhancing, rather than replacing, human judgment.

\subsection{Human-AI Collaboration Framework for Civic Communication}

To incorporate AI into civic communication processes, we must first conceptualize the appropriate relationship between human and automation.
While various frameworks illustrate different conditions of human-AI collaboration~\cite{liWhereAreWe2024, lai2019human}, we adapt Shneiderman's human-centered AI framework as particularly applicable to civic communication contexts~\cite{shneidermanHumanCenteredArtificialIntelligence2020}.
Unlike frameworks that position human and AI control as opposing ends of a one-dimensional spectrum, Shneiderman emphasizes that human control and computer automation exist on separate dimensions, forming a two-dimensional grid of possibilities.

In Shneiderman's framework, systems can exhibit various combinations of human control and computer automation. 
Some systems, like pacemakers, operate with high computer automation but minimal human control, while others like musical instruments provide high human control with limited automation. 
For complex contexts with varying requirements, like the tasks in community engagement, Shneiderman advocates for systems with both high human control and high computer automation (e.g. elevators), describing these as ``reliable, safe, and trustworthy systems'' that augment human capabilities while preserving agency. 

\begin{figure}[h]
  \centering
  \includegraphics[width=\linewidth]{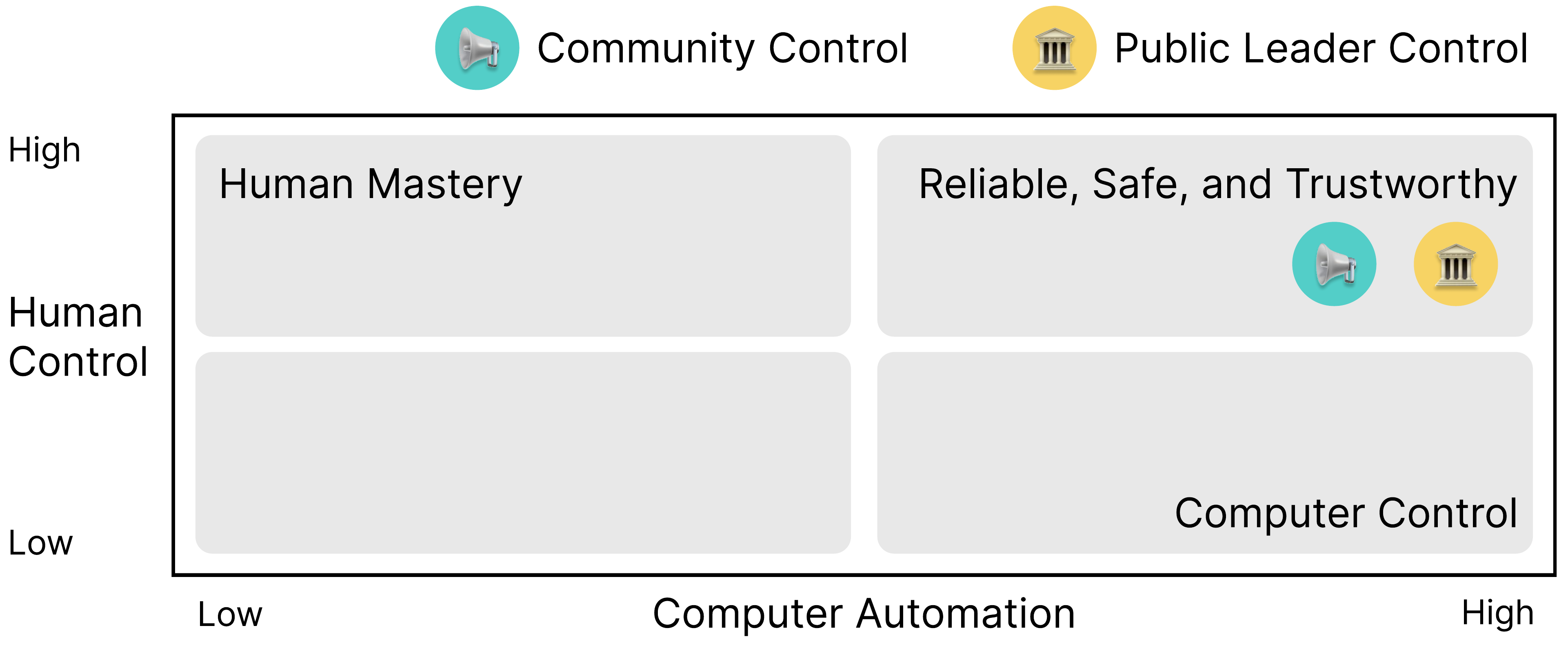}
  \caption{Shneiderman’s human-AI collaboration framework}
  \label{fig:figure2}
\end{figure}

For civic communication specifically, this framework requires further refinement to recognize two distinct types of human control: public leader control and community organization control.
The updated framework is depicted in Fig.~\ref{fig:figure2}.
Public leaders need sufficient control over AI systems to ensure communication aligns with project constraints, while community organizations require control to maintain authentic representation in engagement processes.
Therefore, the optimal quadrant in the adapted framework becomes one with high automation alongside high control for both stakeholder groups, avoiding scenarios where AI increases efficiency at the expense of either group's agency.
By centering both types of human control alongside automation, public leaders and communities can harness AI's capabilities while preserving the human-driven nature of democratic processes.

\subsection{Design Considerations for Human-AI Collaboration}

To implement Shneiderman’s human-AI collaboration framework in civic communication, designers must carefully balance several considerations (i.e. critical engagement, flexibility, transparent boundaries, feedback loops, and reflection) to foster high control for both public leaders and community organizations while leveraging automation benefits. 

A fundamental tension exists between critical engagement with AI outputs and system usability. 
Humans frequently process AI-generated information superficially~\cite{gajosPeopleEngageCognitively2022}, creating vulnerability to hallucinations and biases. 
While interventions like cognitive forcing functions~\cite{bucincaTrustThinkCognitive2021} or speed bumps~\cite{stackpole2024speedbumps} can counter this tendency, they may negatively impact user experience. 
Systems promoting critical engagement must offer clear benefits over quick acceptance, making the evaluation of AI outputs feel natural rather than burdensome.
Furthermore, community engagement efforts vary widely in complexity, stakeholder composition, and goals, requiring AI systems with substantial flexibility and user control.
Mixed-initiative systems highlight the importance of multiple interaction modes that adapt to different user needs and expertise levels~\cite{horvitzPrinciplesMixedinitiativeUser1999}. 
This flexibility includes varied feedback mechanisms~\cite{overneySenseMateAccessibleBeginnerFriendly2024}, while user-driven interactions require choosing when to receive AI assistance and providing means to invoke, refine, or terminate services~\cite{amershiGuidelinesHumanAIInteraction2019}. 
In community engagement, where stakeholders possess varied expertise, customization options can support diverse participation styles~\cite{overney2025coalesce}, increasing inclusivity.
The effectiveness of human-AI collaboration depends on clear definitions of complementary roles and boundaries. 
The different roles connect with the concept of complementary strengths, in which humans and AI bring different expertise to a task, enhancing the performance of human-AI teams compared to human-only or AI-only configurations~\cite{overney2025coalesce, zhang2020effect}.
Boundaries determine how much machine influence is acceptable in a process which aims to surface human opinions~\cite{smallOpportunitiesRisksLLMs2023}, encouraging explicit restrictions on AI in community engagement.
Throughout multi-step processes, humans and AI can dynamically shift roles as creators, assistants, optimizers, and reviewers~\cite{liWhereAreWe2024}.
Systems should make these shifts transparent to help users maintain an ``intentional stance'' toward AI~\cite{dennett1987intentional}, which involves viewing AI as having ``capacities to influence''~\cite{geHowCultureShapes2024} as a rational agent with specific goals and limitations.
Human-AI collaboration requires robust feedback mechanisms enabling iterative communication and continuous learning~\cite{horvitzPrinciplesMixedinitiativeUser1999, amershiGuidelinesHumanAIInteraction2019}. 
Effective systems provide rapid, incremental actions with informative feedback~\cite{shneidermanHumanCenteredArtificialIntelligence2020} and maintain a memory of interactions~\cite{horvitzPrinciplesMixedinitiativeUser1999}. 
While explanations can improve collaboration~\cite{kimHelpMeHelp2023}, they must enhance transparency without creating false confidence~\cite{bucincaTrustThinkCognitive2021}, particularly when AI systems interpret diverse community perspectives.
Beyond feedback loops, civic AI systems should actively promote deeper reflection and deliberation about community input.
Zhang et al. distinguish between ``reflection-in-action'', intuitive reasoning during a task, and ``reflection-on-action'', retrospection after a task~\cite{zhangDeliberatingAIImproving2023}.
AI can support these reflection modes by framing explanations as questions rather than declarations~\cite{danryDonJustTell2023}, prompting users to generate self-explanations that enhance critical thinking. 
After promoting self-reflection, AI systems can be extended to facilitate deliberation or dialogue, helping people share perspectives and reach common understanding despite differences~\cite{zhangDeliberatingAIImproving2023}.
Particularly promising approaches involve AI-support for facilitation among humans with different viewpoints~\cite{korre2025evaluationfacilitationonlinediscussions} or even simulated deliberation or decision-making between AI agents representing diverse perspectives~\cite{du2023improving, gordonJuryLearningIntegrating2022}.
An emphasis on deliberation can create structured opportunities for perspective-taking and perspective-getting, which can reduce ideological divides. 
As Lee et al. observe, participatory AI approaches may help find consensus in polarized contexts, but must be paired with deliberative techniques that bring together diverse parties to enhance constructive communication~\cite{leeWeBuildAIParticipatoryFramework2019}.

%% file: conclusion.tex
\section{Conclusion}

Constructive communication in community engagement requires fostering bidirectional exchanges characterized by understanding, trust, respect, legitimacy, and agency among public leaders and community organizations.
Across different communication pathways, whether between public leaders and community organizations or among community groups themselves, persistent challenges include resource constraints, sensemaking difficulties, engagement fatigue, and barriers to inclusive participation. 
Human-AI collaboration offers promising avenues to address these challenges by streamlining resource-intensive workflows, revealing overlooked perspectives, enhancing mutual understanding, and fostering reflection across social divides. 
However, these potential benefits must be weighed against significant AI risks including accuracy limitations, transparency gaps, bias amplification, and threats to human agency that could undermine the human-centered nature of civic dialogue.
Achieving beneficial human-AI collaboration in community engagement depends on design approaches that maintain a high degree of control in decision-making for both public leaders and community organizations while leveraging automation capabilities.
This requires balancing critical engagement with usability, creating flexible and user-driven interactions, defining transparent roles between humans and AI, building robust feedback loops, and actively encouraging reflection and deliberation.
When designed with these considerations, AI-based civic systems can strengthen the communication pathways in community engagement processes while mitigating common risks that come with technology-enabled participation. 